\documentstyle[aps,pra%
,preprint%
]{revtex}
\begin{document}
\draft
\title{DRY FRICTION IN THE FRENKEL-KONTOROVA-TOMLINSON MODEL:
   \\ DYNAMICAL PROPERTIES}
\author{Michael Weiss\cite{MWadr} and Franz-Josef Elmer}
\address{Institut f\"ur Physik, Universit\"at
   Basel, CH-4056 Basel, Switzerland}
\maketitle
\begin{abstract}
Wearless friction is investigated in a simple mechanical model called
Frenkel-Kontorova-Tomlinson model. We have introduced this model in 
[Phys. Rev. B, {\bf 53}, 7539 (1996)] where the static friction has
already been considered. Here the model is treated for constant
sliding speed. The motion of the internal degrees of freedom is
regular for small sliding velocities or weak interaction between the
sliding surfaces. The regular motion for large velocities is strongly
determined by normal and superharmonic resonance of phonons excited
by the so-called ``washboard wave''. The kinetic friction has maxima
near these resonances. For increasing interaction strength the
regular motion becomes unstable due to parametric resonance leading
to quasistatic and chaotic motion. For sliding velocities beyond
first-order parametric resonance bistability occurs between the 
strongly chaotic motion (fluid sliding state), where friction is
large and a regular motion (solid sliding state), where friction is
weak. The fluid sliding state is mainly determined by the density of
decay channels of $m$ washboard waves into $n$ phonons. This density
describes qualitatively the effectiveness of the energy transfer from
the uniform sliding motion into the microscopic, irregular motion of
the degrees of freedom at the sliding interface.  For a narrow
interval of the sliding velocities we also found enhanced friction
due to coherent motion. In the regime of coherent motion
non-destructive interactions of dark envelope solitons occur.
\end{abstract}

\narrowtext

\section{\protect\label{INT}Introduction}

The sliding of two solid bodies relative to one another is a
non-equilibrium process where the kinetic energy of the uniform
motion is transferred into energy of irregular microscopic motion,
i.e. into heat. This dissipative process leads to friction, called
{\em dry friction\/}, which differs from viscous friction. Since
more than 200~years the phenomenological laws of dry friction
(Coulomb's laws) are well-known and are well established in applied
physics \cite{bow.54}. They state that the friction force is given by
a material parameter (friction coefficient) times the normal force.
The coefficient of static friction (i.e., the force necessary to
start sliding) is larger or equal to the coefficient of the kinetic
friction (i.e., the force necessary for sliding at a constant
velocity).

One cannot expect that Coulomb's law can be derived as rigorously as
the laws of viscous friction. The reason is that the latter one can
be deduced from the well-established theory of near-equilibrium
thermodynamics whereas dry friction works mostly far from thermal
equilibrium. In fact Coulomb's laws oversimplify a very complex
behavior which involves elastomechanical, plastic, and chemical
processes operating on different length and time scales (for a
general overview of the interdisciplinary aspects of dry friction see
Ref.~\onlinecite{sin.92}). Experimentally deviations of Coulomb's law
are often found \cite{sin.92}. For example, the kinetic friction
depends on the sliding velocity. A common pattern is that it first
decreases, then goes through a minimum and finally increases
\cite{hes.94}. In the case of a thin lubrication layer (thickness: a 
few monolayers) one often finds the contrary: The kinetic friction
first increases, then goes through a maximum and finally decreases
\cite{bhu.95}.

In the last ten years the interest in dry friction (i.e., friction
between solids) has been aroused again due to the possibility of
reproducible friction experiments on the mesoscopic and nanoscopic
scale (nanotribology) \cite{sin.92}. There is some hope that for such
less complex systems an understanding is possible by theoretical
modeling \cite{bhu.95}. From the general point of view of
nonequilibrium thermodynamics there is also some hope that it should
be possible to develop a simple model which shows all the basic
phenomenological features of dry friction. Like the Ising model in
the theory of ferromagnetism it is acceptable that such a model is not
realistic in every details. What is more important is that it is on 
the one hand sufficiently complex to be a testing ground of basic 
concepts of the statistical mechanics of dry friction. On the other 
hand it should be simple enough to get analytical results. 

The Frenkel-Kontorova-Tomlinson (FKT) model goes back to an old idea
of Tomlinson which explains Coulomb's laws on the atomic level
\cite{tom.29,mcc.89}. His attempt concerns the most puzzling feature
of dry friction, namely, that the friction force remains finite in
the limit of quasistatic sliding (i.e., sliding velocity going to
zero). It is clear that this behavior cannot be treated, like viscous
friction, as the linear response of a system near thermal
equilibrium. Tomlinson's idea was that friction is caused by mutual
``plucking'' of surface atoms. Each plucked surface atom vibrates.
Plucking is possible only if the sliding brings surface atoms into
metastable states from which they suddenly jump into more stable
ones. The energy difference is turned into kinetic energy which is
totally dissipated due to the emission of some type of bulk or
surface waves (e.g., sound waves) by the plucked atom.  This is the
basic dissipation channel. Tomlinson's mechanism works even in the
limit of quasistatic sliding provided that the barrier between the
metastable and the stable states is sufficiently high and the
temperature sufficiently low.

This simple picture assumes (i) that the relaxation time of a plucked
atom is much smaller than the averaged time between two jumps of the
same atom and (ii) that a vibrating atom does not excite vibrations
of other surface atoms. These assumptions lead to a kinetic friction
that does not depend on the sliding velocity \cite{tom.29,mcc.92}.

The first assumption does not hold if either the vibration after the
jump is not totally dissipated until the next plucking event of the
same atom (weak dissipation) or the jump itself occurs on a time
scale larger than the averaged time between two plucking events of a
single atom (strong dissipation). One can imagine that in the case of
atomically flat surfaces such plucking events occur in a regular
fashion with a frequency (so-called {\em washboard
frequency\/}\cite{sok.84}) which is determined by the sliding
velocity multiplied by the principal wave number of the surface
structure. In the case of weak damping we can therefore expect {\em
resonances\/} which should lead to a velocity-dependent kinetic
friction with resonance peaks. The situation can be modeled by the
independent oscillator or Tomlinson model \cite{mcc.89,hel.94}. It is
a damped harmonic oscillator (eigenfrequency $\omega_0$) interacting
with a sliding surface (velocity $v$) which is described by a
spatially periodic potential with wave number $k$. Two types of
resonances are possible in the Tomlinson model \cite{elm.94}: {\em
Normal resonance\/} where the oscillator is plucked after $n$
oscillations (resonance condition:  $\omega\equiv vk=\omega_0/n$),
and {\em parametric resonance\/} which is an instability where
oscillations are excited due to the modulation of the effective
frequency of the oscillator during sliding (resonance condition:
$\omega=2\omega_0/n$, where $n$ is an integer)\cite{remINT.1}. In the
case of weak interaction these resonances can be calculated
analytically, whereas in the case of strong interaction the motion of
the oscillator becomes chaotic and pronounced hysteresis
occurs\cite{hel.94,elm.94}. The kinetic friction becomes a wild,
multi-valued function of the sliding velocity. This is even the case
after averaging over an ensemble of {\em identical\/}
oscillators\cite{remINT.2}.

The easiest way of taking into account the coupling between surface
atoms is to introduce a linear nearest-neighbor interaction of the
oscillators. This leads to the Frenkel-Kontorova-Tomlinson model
which we have already introduced in Ref.~\onlinecite{wei.96a}.  The
FKT model assumes two rigid sliding bodies where the layer of surface
atoms of one body is modeled by a mechanical particle-spring model
(see Fig.~\ref{f1}). There are two kind of springs, a nearest-neighbor
spring and a spring which couples each particle to the body. The
interaction between the two bodies is described by a spatially
periodic potential.  The dissipative coupling of the layer with the
environment is modeled by damping terms which are proportional to the
relative velocities.  The static properties (ground state,
meta-stable states, static friction) of the one-dimensional FKT model
have already been discussed in Ref.~\onlinecite{wei.96a}. In this
paper we investigate dynamical properties such as the kinetic
friction of the one-dimensional FKT model. A brief summary of
Ref.~\onlinecite{wei.96a} and of this paper is given in
Ref.~\onlinecite{wei.96b}.

The FKT model is far from being realistic in certain aspects. Its
main disadvantages are the following. The assumption of harmonic
particle interaction does not hold in the case of strong distortions.
The elasticity of the bulk is missing which leads to an artificial
gap in the phonon dispersion relation (see Sec.~\ref{HI.L}).

The FKT model shares many similarities with the Burridge-Knopoff (BK)
model\cite{bur.67}, which was proposed as a model of an earthquake
fault. There also a linear, harmonic chain of blocks attached via
springs at a rigid plate is assumed. But the interaction with the
other surface is modeled by phenomenological force describing a dry
friction law.  Mostly velocity-weakening laws are used. That is, the
friction force decreases with the sliding velocity. The dynamical
properties of the BK model have been studied extensively in the
literature \cite{car.89,car.91,vas.92,des.93,%
sch.93,des.94,esp.94,sar.95,des.96}.

The paper is organized as follows. In Sec.~\ref{DEF} the FKT model 
and the kinetic friction $F_K$ are defined. The behavior for small
sliding velocities is treated in Sec.~\ref{LOW}. Here mostly
stick-slip motion of individual particles occurs. At higher
velocities excitations of phonons become important. The consequence
of several kinds of resonances is discussed in Sec.~\ref{HI}. In the
last section we compare our results with other models, especially
with the BK model.

\section{\protect\label{DEF}Definitions}

The Frenkel-Kontorova-Tomlinson (FKT) model has already be introduced
in Ref.~\onlinecite{wei.96a}. Thus the key features are described
only briefly. Here we will concentrate more on the dynamical
properties of the model which have not been investigated in
Ref.~\onlinecite{wei.96a}. Also a definition of kinetic friction
$F_K$ is given.

\subsection{\protect\label{DEFM}The FKT model}

A linear chain of $N$ particles coupled harmonically forms the
backbone of the FKT model (see Fig.~\ref{f1}). The particles can move
only parallel to the sliding direction. Thus the model is completely
one-dimensional.  All particles have the same mass which defines the
mass unit. Also the nearest-neighbor interaction is assumed to be
uniform which defines the force unit. The chain is the interface
between two rigid sliding bodies which are modeled in two very
different ways. The interaction with the lower body is described by a
spatially periodic potential.  Its periodicity defines the unit of
the length scale. The interaction with the upper body is assumed to
be harmonic. That is, each particle is coupled via a spring to the
upper body. All springs are identical and they are attached with
equal spacing at the upper body.  The boundary conditions are assumed
to be periodic.

In this paper the motion of the chain is investigated for the case of
sliding the upper body against the lower one at constant velocity
$v$. This externally constrained sliding leads to vibrations within
the chain.  These vibrations excite waves into the bulk of the
sliding bodies which carries energy away from the sliding interface.
In the FKT model the degrees of freedom which are responsible for
these waves are not explicitly present. We include their effect by
introducing two phenomenological damping forces which are
proportional to the particle velocity measured relative to the lower
and the upper body, respectively.  In this paper we investigate the
FKT model in the limit of zero temperature where fluctuation forces
disappear.  Below we will see that the kinetic friction is directly
given by the rate of energy which is transferred into these basic
dissipation channels.

The equation of motion of the FKT model reads
\begin{eqnarray}
  \ddot\xi_j+\gamma_L(v+\dot\xi_j)+\gamma_U\dot\xi_j&=&\xi_{j+1}+
  \xi_{j-1}-(2+\kappa)\xi_j+\nonumber\\&&+b\sin 2\pi(vt+cj+\xi_j),
  \label{DEF.eqm}
\end{eqnarray}
where $\xi_j$ is the position of particle $j$ relative to its
equilibrium value given by the upper body, $c$ is ratio of the
lattice constant of the upper body to that of the lower body, $b$ and
$\kappa$ are the strengths of interaction with the lower and upper
body, respectively, and $\gamma_L$ and $\gamma_U$ are the damping
constants of the particle motion relative to the lower and upper
body, respectively. The boundary condition is
\begin{equation}
  \xi_{j+N}=\xi_j,
  \label{DEF.bc}
\end{equation}
which requires
\begin{equation}
  c=\frac{M}{N},\quad\mbox{with $M$ integer}.
  \label{DEF.c}
\end{equation}
The transformation 
\begin{equation}
   t \to t+\gamma_L/\kappa,\quad\xi_j \to \xi_j-\gamma_Lv/\kappa
   \label{DEF.t2}
\end{equation}
turns (\ref{DEF.eqm}) into
\begin{eqnarray}
  \ddot\xi_j+\gamma\dot\xi_j&=&\xi_{j+1}+
  \xi_{j-1}-(2+\kappa)\xi_j+\nonumber\\&&+b\sin 2\pi(vt+cj+\xi_j),
  \label{DEF.eqm2}
\end{eqnarray}
with
\begin{equation}
  \gamma=\gamma_L+\gamma_U.
  \label{DEF.gamma}
\end{equation}

In addition to the symmetries of the FKT model already mentioned in
Ref.~\onlinecite{wei.96a}, eq.~(\ref{DEF.eqm}) is invariant under 
translation in time by an integer multiple of the inverse
sliding velocity. Below we will see that all solutions which do not
break this symmetry are defined by a single one-parameter function,
the dynamical hull function, which is similar to the static hull
function \cite{wei.96a}.

For the results presented in the following sections, the equation of 
motion (\ref{DEF.eqm2}) is often integrated numerically. We have done
this by using a second-order Runge-Kutta scheme which in the
non-damping case is identical with the Verlet algorithm
\cite{hee.86}. For the time step we have chosen $0.01$. This is
sufficient because all simulations are done for $\kappa=1$ and
$v<1.5$. The corresponding fastest time scale is $2\pi/\sqrt{5}$.

\subsection{\protect\label{DEFK}The kinetic friction}

The kinetic friction $F_K$ is the temporal average of the force
necessary to keep the body sliding at a constant velocity $v$. It is 
given by the sum of the forces between the particles and the upper
body \cite{wei.96a},
\begin{equation}
   F_K=-\lim_{T\to\infty}\frac{\kappa}{T}\int_0^T
   \sum_{j=1}^N\xi_j(t)\,dt.
  \label{DEF.FK1}
\end{equation}

There is a second equivalent definition which clearly shows the
dissipative nature of the kinetic friction. The idea is that after
the decay of transients the energy flows through the system at a
constant rate.  That is, the energy which is put into the system per
unit time by sliding is on average totally dissipated. Thus the total
energy (i.e., the kinetic and the potential energy) of the chain is
on average constant. With the further assumption that the motion of
the $\xi$'s is bounded we get
\begin{equation}
   F_K=\gamma_LvN+\lim_{T\to\infty}\frac{\gamma}{vT}\int_0^T
   \sum_{j=1}^N\dot\xi^2_j(t)\,dt.
  \label{DEF.FK2}
\end{equation}

From the last definition we clearly see that the kinetic friction is
proportional to the {\em rate\/} of energy transfer from the motion
of the interface particles into some types of bulk waves. This is a
general principle: The kinetic friction is determined by the various
rates of energy transfer starting from the uniform macroscopic motion
of the sliding bodies and ending in some dissipation channels. For
the FKT model we are mainly interested in the efficiency of the
transfer from the macroscopic sliding into microscopic motion of the
interface particles. In section~\ref{HI} we will see that this
efficiency is rather high if phonons are excited resonantly. That is,
the kinetic friction as a function of the sliding velocity shows
resonance peaks.

In the following sections we always discuss the kinetic friction
without the first term of definition (\ref{DEF.FK2}). That is, in any
plot of $F_K$ versus $v$ one should add $\gamma_LvN$. We will see
that for very large velocities this term dominates the second one.

\section{\protect\label{LOW}The low-velocity regime}

In this section we investigate the FKT model for very small sliding
velocities. In fact we are treating the quasistatic limit $v\to 0$.
In Ref.~\onlinecite{wei.96a} we have already discussed the
disappearance of the kinetic friction in this limit if $b$ is less
than some critical value $b_c^K$. Here we investigate $F_K$ as a
function of $v$ for $b$ below and above this threshold.

\subsection{\protect\label{LOW0}\protect\lowercase{$b<b_c^{\protect
   \uppercase{K}}$}}

In Ref.~\onlinecite{wei.96a} we have shown that for $b<b_c^K$ the
stationary values of the $\xi$'s follow the quasistatic sliding
adiabatically. Furthermore, there is only one stationary state which
can be described by the static hull function $g_S$ [defined by
Eq.~(10) in Ref.~\onlinecite{wei.96a} which is identical with
(\ref{HI.heqm}) for $v=0$]. Therefore, the solution of the equation
of motion (\ref{DEF.eqm2}) in the quasistatic limit reads
\begin{equation}
  \xi_j(t)=g_S(vt+cj)\quad{\rm with}\quad g_S(x+1)=g_S.
  \label{LOW.gs}
\end{equation}
Using (\ref{DEF.FK2}) we immediately see that the kinetic friction
behaves viciousely, i.e.,
\begin{equation}
  F_K=\gamma_{\rm eff}vN,
  \label{LOW.FK1}
\end{equation}
with an effective damping constant
\begin{equation}
  \gamma_{\rm eff}=\gamma_L+\gamma\int_0^1
      \left(\frac{dg_S}{dx}\right)^2dx.
  \label{LOW.gamma}
\end{equation}
In the Tomlinson limit (i.e., $\kappa,b\to\infty$) and for $c=1$ an
analytic expression can be obtained (see appendix~\ref{AP1}). In the
limit $b\to b_c^K$ the effective damping constant diverges like
\begin{equation}
  \gamma_{\rm eff}\sim (b_c^K-b)^{-1/2}.
  \label{LOW.ga.scale}
\end{equation}

\subsection{\protect\label{LOW1}\protect\lowercase{$b>b_c^{\protect
   \uppercase{K}}$}}

For $b>b_c^K$ additional metastable states appear \cite{remLOW.1}.
Due to quasistatic sliding these states eventually annihilate with an
unstable counterpart in a saddle-node bifurcation \cite{wei.96a}.  At
those points the chain has to find a new metastable configuration.
That is, the particles of the chain no longer move adiabatically.
They start to move with finite velocities, and after a transient they
reach a new metastable state. The dissipated energy is the difference
in potential energy before and after the reconfiguration event. This
is the generalization of Tomlinson's friction mechanism. The kinetic
friction in the quasistatic limit is therefore the averaged energy
drop per jump $\langle\Delta V\rangle$ divided by the averaged
distance $\langle\Delta x_B\rangle$ between the relative positions of
the sliding bodies between two subsequent reconfiguration events:
\begin{equation}
 F_K(v\to 0)=\frac{\langle\Delta V\rangle}{\langle\Delta x_B\rangle}.
 \label{LOW.FK0}
\end{equation}

In the overdamped limit, i.e., $\gamma\to\infty$, the system selects
via steepest descent the nearest possible metastable state on the
energy surface. Starting with an arbitrary metastable state the
system will reach after several jumps a state which can be obtained
by adiabatic deformation from the ground state of the undriven FKT
model \cite{gya.94}. Then this state always reappears after each
jump. The reason for this selection rule is closely related to the
fact that the ground state has the highest depinning force of all
metastable states \cite{wei.96a}. The sequence of reconfiguration
events becomes therefore regular. We assume that $c$ is the ratio of
two coprime integers $P$ and $Q$ (i.e., $c=P/Q$) where $Q$ is a
divisor of $N$. In each reconfiguration event $N/Q$ equally spaced
particles jump into the neighboring potential wells. Each jumping
particle together with the neighboring particles dissipates $\Delta
V_0$. Thus the totally dissipated energy is $\Delta V_0N/Q$. The
distance between two events is $1/Q$. Thus
\begin{equation}
  F_K(v\to 0)=\Delta V_0N.
  \label{LOW.FK0.o}
\end{equation}
Near the onset of friction $F_K$ scales like
\begin{equation}
  \frac{F_K(v\to 0)}{N}\sim(b-b_c^K)^2.
  \label{LOW.FK0.c}
\end{equation}
The leading orders of this expansion are calculated in
Appendix~\ref{AP1} for $c=1$ and for the Tomlinson limit. For 
$b\gg\kappa\gg 1$ we get (see appendix~\ref{AP1})
\begin{equation}
  \frac{F_K(v\to 0)}{N}=b\left(1-\frac{\kappa}{2b}+\frac{\sqrt{2}}
   {3\pi}\left(\frac{\kappa}{b}\right)^{\frac{3}{2}}
   +\cdots\right).
  \label{LOW.FK0.infty}
\end{equation}
In the limit $\kappa/b\to 0$ the kinetic friction is approaching the
static friction $F_S=F_S^{max}\equiv bN$. 

Figure~\ref{f.low} shows for several values of $c$ the static
friction $F_S$ and the kinetic friction $F_K(v\to 0)$ which are
obtained numerically by solving the delay equation of the hull
function $g_S$. First of all, we notice the excellent agreement
between analytical and numerical results for $c=1$.  Furthermore, we
observe that the kinetic friction is approaching the static friction
from below not only for $b\to\infty$ but also for $Q\to\infty$ (i.e.,
for decreasing commensurability of $c$). In fact the difference
scales like $1/Q$ for $Q\gg 1$. This can be understood by the
following argument. The jumps of the $N/Q$ particles represent a kind
of microslip leading to typical stick-slip motion with a sawtooth
curve of the lateral force $F=-\kappa\sum_j\xi_j$ as a function of
the relative position between the sliding bodies. The maximum of this
curve defines $F_S$ whereas the kinetic friction is the average in
accordance with definition (\ref{DEF.FK1}). Thus $F_S-F_K(v\to 0)$ is
given by half the height of the jumps which is proportional to $1/Q$
because in each jump $N/Q$ particles are jumping over a distance
which is independent of $Q$ for $Q\gg 1$. The jump is roughly
proportional to the largest gap in the hull function $g_S$ (see
Figs.~2 in Ref.~\onlinecite{wei.96a}). 

If we take inertia into account, i.e., $\gamma$ finite, the system
does not necessarily select the nearest possible minima of the energy
landscape. That is, the particle may jump over several potential
wells before being captured. This has two consequences: (i) It lowers
the kinetic friction which can become zero in the extreme case of
$\gamma\to 0$. (ii) It makes the motion irregular. That is, the
states between the reconfiguration events are no longer equivalent to
the ground state, and $\Delta V$ and $\Delta x_B$ are fluctuating
properties.

This irregular behavior is similar to the irregular motion in the BK
model \cite{car.91}. For example, as in the BK model the FKT model
shows a power law distribution of $\Delta V$ up to a certain cutoff
value \cite{elm.96}. This cutoff monotonically decreases with
increasing $\kappa$. For $\kappa\gtrsim 1$ the power law disappears
and the distribution is exponential \cite{elm.96,des.96}. The power
law behavior has also fed the speculation that dry friction may be a
self-organized critical process. This subject is not addressed here.
The interested reader is referred to Ref.~\onlinecite{elm.96} where
the FKT model is discussed in this context. 

The dynamical behavior of the chain does not change qualitatively if
the sliding velocity is finite but still small. There is again a
regular sequence of reconfiguration events where $N/Q$ equally spaced
particles jump. It is expected that $F_K(v)$ increases with $v$
because the critical slowing down of the system near a saddle-node
bifurcation delays the reconfiguration event. Thus the energy drop
$\Delta V$ is larger than in the quasistatic limit because at the
beginning of the reconfiguration event the strings connecting the
particles with the upper body are stretched more.  From simulations
we find that $dF_K/dv|_{v\to 0}$ is indeed positive (see
Figs.~\ref{f.lows} and \ref{f.fk}). We also see that after reaching a
maximum $F_K$ decreases.  This can be understood in the following
way. In the simulations shown in Figs.~\ref{f.lows} and \ref{f.fk}
the motion is not overdamped. Therefore the particles oscillate after
jumping. If the sliding is sufficiently fast they are still
oscillating at the next jumping event.  Thus they explore the phase
space and may therefore overcome the barrier before it disappears in
the saddle-node bifurcation mentioned above. For increasing sliding
velocity the delay caused by critical slowing down is more and more
compensated by this mechanism.  It leads to a local maximum of
$F_K(v)$ at a velocity which scales like $\gamma$. This is roughly
confirmed by the friction curves in Fig.~\ref{f.lows}.

\section{\protect\label{HI}The high velocity regime}

In the previous section we have investigated the FKT model in a
regime where Tomlinson's basic dissipation mechanism works, namely,
that the energy put into the system is released and subsequently
dissipated in recognizable reconfiguration events. For increasing
sliding velocity the average time between successive reconfiguration
events will eventually become smaller than the average time scale on
which such events take place. Reconfiguration events overlap and are
no longer recognizable. Thus we have entered a new regime where the
particles are always in motion. In this regime the friction may be
strongly enhanced due to {\em resonances of waves\/}, i.e., phonons. 

In this section the consequences of these resonances are investigated
on various levels. First we treat the linearized equation of motion.
In a second step we calculate numerically all regular solutions which
do not break the discrete temporal translation symmetry of the
equation of motion. Then irregular solutions are investigated. At the
end of this section we report on solitary behavior.

\subsection{\protect\label{HI.L}The basic resonances}

The external potential describing the hard surface of the lower body
makes the equation of motion (\ref{DEF.eqm2}) nonlinear. Therefore it
is very hard to find analytical solutions. For a weak potential,
i.e., $b\ll 1$, and sufficiently strong dissipation one can expect
that the oscillations of the particles are small. Thus we can
linearize the equation of motion,
\begin{eqnarray}
  \ddot{\xi}_j+\gamma\dot{\xi}_j&=&\xi_{j+1}-(2+\kappa)\xi_j+
   \xi_{j-1}\nonumber\\&&+b\sin 2\pi(vt+cj)
   \nonumber\\&&+2\pi b\cos 2\pi(vt+cj)\cdot\xi_j.
  \label{HI.leqm}
\end{eqnarray}
Both terms on the right-hand side which are proportional to $b$ can
be considered as driving terms. These driving terms are periodic in
$t$ and $j$ with frequency $2\pi v$ and wave number $2\pi c$. We call
this driving wave the ``washboard wave''. It leads to different types
of resonances. The first driving term is responsible for the {\em
main resonance\/}. The last term leads to {\em parametric
resonance\/}.  Both terms are responsible for {\em superharmonic
resonance\/}.

The washboard wave can resonantely excite phonons (i.e., sound waves)
which are solutions of (\ref{HI.leqm}) for $\gamma=b=0$. They have
the form 
\begin{displaymath}
  e^{i[k_mj\pm\omega(k_m)t]},
\end{displaymath}
with the dispersion relation 
\begin{equation}
  \omega(k)=\sqrt{\kappa+4\sin^2\frac{k}{2}}.
  \label{HI.disp}
\end{equation}
Because of periodic boundaries, $k$ has to be discrete
\begin{equation}
  k=\frac{2\pi}{N}m,\ m=0,\ldots,N-1.
  \label{HI.k}
\end{equation}

\subsubsection{\protect\label{HI.L.mr}The main resonance}

We drop the last term of (\ref{HI.leqm}). Such linearized equations have
already been studied in many other models \cite{sok.90,per.95,per.96}.
They can be easily solved by the Ansatz
\begin{equation}
  \xi_j=A\cos[2\pi(vt+cj)+\phi]
  \label{HI.xi}
\end{equation}
which leads to a typical resonance line
\begin{equation}
 A=\frac{b}{\sqrt{[\omega^2(2\pi c)-(2\pi v)^2]^2+(2\pi v\gamma)^2}}.
  \label{HI.A}
\end{equation}
Note that (\ref{HI.xi}) is the solution in the limit $t\to\infty$
where all other solutions have died out because of $\gamma>0$. 

In order to calculate from this solution the kinetic friction one has
to be aware that for the linearized equation of motion
(\ref{HI.leqm}) the two definitions (\ref{DEF.FK1}) and
(\ref{DEF.FK2}) are no longer equivalent. The first definition
(\ref{HI.xi}) yields zero friction whereas the second definition
leads to \cite{remHI.1}
\begin{equation}
  \frac{F_K(v)}{N}=\frac{2\pi^2\gamma vb^2}{[\omega^2(2\pi c)-
   (2\pi v)^2]^2+(2\pi v\gamma)^2}.
  \label{HI.FKmr}
\end{equation}
This result is {\em quadratic\/} in $b$ because (\ref{DEF.FK2}) is
quadratic in $\xi$. To get the same result with (\ref{DEF.FK1}) one
has to calculate $\xi_j$ up to second order in $b$ (see
appendix~\ref{AP2}).

For $v\to 0$ the kinetic friction (\ref{HI.FKmr}) goes to zero. Thus
we have the frictionless case $b<b_c^K$ from Sec.~\ref{LOW0}. For
$c=1$ we recover in leading order of $b$ the exact result
(\ref{AP1.gamma}). The
solution (\ref{HI.FKmr}) increases with $v$, reaching the maximum
\begin{equation}
  \frac{F_K}{N}=\frac{b^2}{2\gamma v_1}+{\cal O}(\gamma)
  \label{HI.FKmr.max}
\end{equation}
at
\begin{equation}
  v=v^S_1\equiv\frac{\omega(2\pi c)}{2\pi}+{\cal O}(\gamma^2)
  \label{HI.vmr.max}
\end{equation}
and decreases to zero like $1/v^3$ for $v\to\infty$. The resonance
peak occurs at that washboard frequency $2\pi v$ which corresponds to
the frequency of the phonon having the same wave vector as the
washboard wave.

Qualitatively the same behavior of the kinetic friction as a function
of the sliding velocity is often found in dry friction experiments
between atomically flat surfaces with a few monolayers of
lubricating molecules in between \cite{bhu.95}. Usually such friction
curves are interpreted as a dynamic phase transition from liquid-like
behavior at small velocities to a solid-like state at high
velocities. At the maximum of the kinetic friction the lubrication
layer is assumed to be in an amorphous state \cite{bhu.95,yos.93}. 

We can give an estimate of the validity of the kinetic friction
(\ref{HI.FKmr}). The linearized equation of motion works fine as long
as $|\xi_j|\ll 1$. Using (\ref{HI.A}) we find
\begin{equation}
  b^2\ll [\omega^2(2\pi c)-(2\pi v)^2]^2+(2\pi v\gamma)^2.
  \label{HI.valmr}
\end{equation}
The approximation (\ref{HI.FKmr}) first fails at the resonance peak. 
Thus (\ref{HI.FKmr.max}) is a good approximation as long as $b\ll
\omega(2\pi c)\gamma$. For sliding velocities much larger than
$\sqrt{b}/(2\pi)$ the approximation is always good because the
particles make only tiny oscillations around their equilibrium
positions. We call this state the {\em solid-sliding state\/}.

\subsubsection{\protect\label{HI.L.sr}Superharmonic resonances}

The last term in (\ref{HI.leqm}) is responsible for resonances of 
higher harmonics. For example, the solution (\ref{HI.xi}) would lead
to a term of the form
\begin{displaymath}
  e^{4\pi i(vt+cj)}.
\end{displaymath}
It is therefore natural to expand $\xi_j$ into a Fourier series:
\begin{equation}
  \xi_j=\sum_{m=-\infty}^\infty A_me^{2\pi im(vt+cj)},\quad 
   A_{-m}=A_m^*.
  \label{HI.xij}
\end{equation}
The linearized equation of motion turns into a set of linear 
algebraic equations for the Fourier coefficients $A_m$:
\begin{equation}
  L_mA_m-\pi b(A_{m-1}+A_{m+1})=\frac{ib}{2}\left(\delta_{m,-1}
   -\delta_{m,1}\right),
  \label{HI.Am}
\end{equation}
where $\delta_{i,j}$ is the Kronecker symbol and
\begin{equation}
  L_m\equiv\omega^2(2\pi cm)-(2\pi vm)^2+2\pi vm\gamma\,i.
  \label{HI.Lm}
\end{equation}
For small values of $b$, (\ref{HI.Am}) can be solved perturbatively
(see appendix~\ref{AP2}). In order to express $F_K$ in terms of the
Fourier components $A_m$, we use again definition (\ref{DEF.FK2}) of
the kinetic friction \cite{remHI.1}:
\begin{equation}
  \frac{F_K}{N}=8\pi^2\gamma v\sum_{m=1}^\infty |mA_m|^2.
  \label{HI.FKF}
\end{equation}

Because $A_m$ is proportional to $1/L_m$ we expect a peak in $F_K(v)$
near the {\em superharmonic resonance velocity\/}
\begin{equation}
  v^S_m\equiv \frac{\omega(2\pi cm)}{2\pi m}.
  \label{HI.vms}
\end{equation}
This justifies an approximation of $F_K$ which takes only
the leading term in the power series of $A_m$ into account:
\begin{equation}
  \frac{F_K}{N}\approx 2v\gamma\sum_{m=1}^\infty m^2\,\prod_{j=1}^m
   \left|\frac{\pi b}{L_j}\right|^2.
  \label{HI.FKapp}
\end{equation}
This is not a correct expansion of $F_K$ in powers of $b$ but it
reproduces the superharmonic resonance peaks qualitatively very well
as is shown in figure~\ref{f.res}(a). It also gives the right scaling
of the height of the resonance peaks, namely, $b^{2m}$ for the
superharmonic resonance of order $m$.

A special case of superharmonic resonance is $m=1$ which is the 
main resonances. It can be shown that
\begin{equation}
  v^S_1>v^S_m,\quad{\rm for}\quad m>1.
  \label{HI.sr.rel}
\end{equation}
Thus the superharmonic resonance peaks occur between zero and
$v^S_1$. But they do not order sequentially (i.e., $v^S_n<v^S_m$
for some $n>m$) even though $v^S_m\to 0$ for $m\to\infty$.

\subsubsection{\protect\label{HI.L.pr}Parametric resonances}

The last term in the linearized equation of motion (\ref{HI.leqm}) is
also responsible for parametric resonance of phonons. Parametric
resonance is an instability phenomenon. The prototype is a pendulum
with a vertically oscillating support \cite{ll.76}. The linearized
equation of motion is the well-known Mathieu equation \cite{han.65}
which is a (damped) oscillator with eigenfrequency $\omega_0$. Energy
can be put into the oscillator due to periodic modulation (with
frequency $\omega$) of a parameter which determines $\omega_0$. If
the oscillator is damped this energy will dissipate. 

The behavior of such a parametrically driven oscillator is very
different from an oscillator which is driven by an additive force.
In the latter case the response is linear whereas in the first case
it is strongly nonlinear. The linear relationship between the
amplitude of the oscillation and the strength of an additive force is
caused by the fact that the driving force is {\em independent\/} of
the amplitude of the oscillation whereas in a parametrically driven
oscillator the force is proportional to the amplitude. This leads to
a positive feedback mechanism. For a small proportionality constant
(i.e., strength of driving) more energy is dissipated than is gained
from driving and the oscillation dies out. But if the strength of
driving exceeds a certain threshold, more energy is gained than is
dissipated and the amplitude increases. In a linear equation like the
Mathieu equation such solutions go to infinity. But in real systems
like the pendulum, nonlinearities will eventually saturate this
growth.

The threshold for parametric resonance is a function on the frequency
ratio $\omega_0/\omega$. It has relative minima near the {\em
parametric resonance condition\/}
\begin{equation}
  \omega=\frac{2\omega_0}{m},\quad m=1,2,\ldots
  \label{HI.pr.con}
\end{equation}
where $m$ denotes the order.  The main difference between a
parametrically driven oscillator and the FKT model is that in the
latter case there are several oscillators (i.e. phonons), and the
driving is spatially non-uniform because it is a wave. In this case
the parametric resonance condition reads (see appendix~\ref{AP2})
\begin{equation}
  v^P_m(q)=\frac{\omega(\pi c m+q)+\omega(\pi c m-q)}{2\pi m},\quad
   m=1,2,\ldots
  \label{HI.pr.vn}
\end{equation}
This condition can be intuitively understood by the following
picture: The linear chain is driven by the washboard wave which has
``momentum'' (i.e., wave number) $2\pi c$ and ``energy'' (i.e., 
frequency) $2\pi v$. If the driving is sufficiently strong, phonons
with wave number $k_\pm=\pi cm\pm q$ are excited parametrically by
$m$ washboard waves. The condition (\ref{HI.pr.vn}) follows from the
conservation laws of momentum and energy. It depends on a continuous 
parameter $q$ contrary to the resonance condition (\ref{HI.vms}).  Of
course the periodic boundary condition (\ref{DEF.bc}) restrict the
value of $q$ to $N$ different values corresponding to $N$ pairs of
phonons. For $c=1$ the driving is uniform and (\ref{HI.pr.vn})
becomes identical to (\ref{HI.pr.con}) where $\omega_0$ is the
frequency of the pair of phonons with the same wave number but
propagating in opposite directions and forming a standing wave. For
weak damping the threshold for the $m$-th order parametric resonance
is proportional to $\gamma^{1/m}$ (see in appendix~\ref{AP2}).

The maxima and minima of $v^P_m(q)$ define resonance intervals which
successively appear for increasing $b$ starting with the first-order
parametric resonance. There is a relation between these instability
intervals and the superharmonic resonances:
\begin{equation}
  v^P_m\ge v^S_m.
  \label{HI.prsr}
\end{equation}
Thus the first-order parametric resonance occurs for sliding
velocities larger than the velocity $v^S_1$ of the main resonance.
For $\kappa\to 0$ the minimum of $v_m^P$ is given by $v^S_m$.  It is
possible that a superharmonic resonance peak sits in a parametric
resonance interval, i.e., $\min v^P_m\le v^S_n\le\max v^P_m$ for
$n\not=m>1$. Also overlaps of instability intervals are possible.

For sliding velocities in an instability interval, the amplitudes of 
two phonons which are selected by $v=v_m^P$ increases exponentially.
This exponential increase is saturated by the nonlinear terms omitted
in (\ref{HI.leqm}). Their effects will be discussed in the following
subsections. The increase and saturation of the amplitudes also occur
in $F_K$. We expect that parametric resonance does not lead to a peak
in $F_K(v)$ but to something which looks like a more or less flat
table mountain.  Furthermore, we expect quasiperiodic motion because
the phonon frequencies $\omega(\pi cm+q)$ and $\omega(\pi cm-q)$ are
in general incommensurate except for $c=1$.

\subsection{\protect\label{HI.P}Periodic and quasiperiodic solutions}

From section~\ref{DEFM} we know that the nonlinear equation of motion
(\ref{DEF.eqm2}) is invariant under discrete translation in time.
Solutions with the same symmetry should therefore exist. They can
be described by a one-parameter function called {\em dynamical hull
function\/} $g_D$:
\begin{equation}
  \xi_j(t)=g_D(vt+cj),\quad{\rm with}\quad g_D(x+1)=g_D(x).
  \label{HI.hull}
\end{equation}
The dynamical hull function is very similar to the static hull
function $g_S$ introduced in Ref.~\onlinecite{wei.96a}. In fact for
$v\to 0$ the dynamical hull function becomes identical to the static
one. It is a solution of 
\begin{eqnarray}
  v^2g_D''(x)+\gamma vg_D'(x)&=&g_D(x-c)+g_D(x+c)-\nonumber\\
   &&-(2+\kappa)g_D(x)+\nonumber\\&&+b\sin 2\pi[x+g_D(x)].
  \label{HI.heqm}
\end{eqnarray}
With the Fourier ansatz 
\begin{equation}
  g_D(x)=\sum_{m=-\infty}^\infty A_me^{2\pi imx},\quad A_{-m}=A_m^*.
  \label{HI.gD}
\end{equation}
this differential-delay equation turns into a set of nonlinear
algebraic equations:
\begin{equation}
  L_mA_m=b\int_0^1\!\sin\left(2\pi x+2\pi\sum_{l=-\infty}^\infty A_l
   e^{2\pi ilx}\right)e^{-2\pi imx}dx.
  \label{HI.gDm}
\end{equation}
We solve this equation numerically by truncating the number of
Fourier coefficients. Figure~\ref{f.res}(b) shows the result for an
increasing sequence of values of $b$.

We see that the dynamical hull function is not uniquely defined if
$b$ exceeds some threshold. The resonance peaks starting with the
main resonance develop loops leading to bistability and hystereses.
This is caused by the fact that the oscillation amplitude of the
particle becomes of the same order as the periodicity of the external
potential. These loops are also found in a single Tomlinson
oscillator \cite{hel.94,elm.94}.

The nonlinear terms lead also to a reduction of the height of the
resonance peaks compared to what is expected from the linear theory
[see e.g. Fig.~\ref{f.res}(a)].  Because the peak height in the
linear theory is proportional to the inverse damping constant [see
(\ref{HI.FKmr.max})], we interpret this result as an increase of the
effective damping constant. The reason for this increase is the
opening of additional channels of energy dissipation due to nonlinear
phonon coupling.

In order to investigate the stability of the solution described by
the dynamical hull function, one has to study the equation of motion
linearized around (\ref{HI.hull}). That is, we consider only those
terms which are linear in $\delta\xi_j\equiv\xi_j-g_D(vt+cj)$.  We
get 
\begin{eqnarray}
  \lefteqn{\ddot{\delta\xi}_j+v\dot{\delta\xi}_j=\delta\xi_{j+1}
   +\delta\xi_{j+1}-(2+\kappa)\delta\xi_j+}\nonumber\\&&\hspace{10mm}
   +2\pi b\cos\bigl(2\pi[vt+cj+g_D(vt+cj)]\bigr)\,\delta\xi_j.
  \label{HI.leqm2}
\end{eqnarray}
With the help of the well-known Floquet-Bloch ansatz for
$\delta\xi_j$ (see appendix~\ref{AP2}) we solve this equation 
numerically with truncated Fourier expansions of $g_D$ and
$\delta\xi_j$.

There are two types of instabilities of the hull function. The first
type is caused by the bistability of the resonance loop because there
is always a separating unstable solution between the stable ones
(called a saddle in the terminology of bifurcation theory). The other
instability type is caused by parametric resonance already discussed
for the linearized equation of motion (\ref{HI.leqm}) in
Sec.~\ref{HI.L.pr}. From the stability analysis in appendix~\ref{AP2}
we know that the most destabilizing disturbance contains two
generically incommensurable frequencies ($2\pi v$ and $\Omega$). This
is confirmed numerically for (\ref{HI.leqm2}). At these instabilities
the system selects a state which leads most often to an enhanced
kinetic friction. An example for this enhancement is shown in
Fig.~\ref{f.res}(c). The most prominent enhancement occurs for
first-order parametric resonance. 

Figure~\ref{f.quasi}(a) shows the quasiperiodic motion expected from
Sec.~\ref{HI.L.pr}. It shows a stroboscopic map of the lateral force
in time steps given the ratio of the lattice constant of the external
potential divided by the velocity. The lateral force is given by
\begin{displaymath}
 f_n\equiv -\frac{\kappa}{N}\sum_{j=1}^N\xi_j(n/v).
\end{displaymath}
The dynamical hull function corresponds to a fixed point of the map
\begin{displaymath}
  f_{n+1}=T(f_n).
\end{displaymath}
Quasiperiodic motion leads to a closed one-dimensional attractor of
the map. For $c=1$ the motion is periodic leading
to a limit cycle of period two.

\subsection{\protect\label{HI.I}Irregular behavior}

For increasing driving strength $b$ the quasiperiodic solutions
caused by parametric resonance becomes unstable and the motion become
more and more irregular and chaotic [see Fig.~\ref{f.quasi}(b)]. It
is presumably spatio-temporally chaotic as in the Frenkel-Kontorova
model \cite{str.96}. That is, the number of positive Lyapunov
exponents is an extensive quantity \cite{cro.93}, i.e., it is
proportional to $N$. We call this strongly mixing state a {\em
fluid-sliding state\/} even though the harmonic interaction in the
model excludes on average changes in the particle order.
Nevertheless the state is fluid-like because the amplitudes of the
particle motion are of the same order as the lattice constant $c$.

The kinetic friction $F_K$ from simulations are shown in 
Figure~\ref{f.fk}. In each simulation the sliding velocity was
increased in tiny steps from zero up to a certain value and then
decreased back to zero. Beyond first-order parametric resonance
(i.e., $v>\max v_1^P$) large hysteresis loops occur leading to
bistability between fluid-sliding states and the solid-sliding state.

Characteristic peaks appear in $F_K(v)$ in the regime of
fluid-sliding states beyond the main resonance. What is the nature of
these peaks?  Looking at the spatially and temporally resolved
dynamics we were not able to see any difference between the dynamics
inside and outside the peaks. The motion is strongly irregular. The
only exceptions are the large peaks at $v=1.06$ and $v=0.95$ for
$c=144/233$ and $c=1$, respectively. Here coherent motion appears
which is investigated in the following subsection.  

The spatio-temporal Fourier transformation of $\xi_j(t)$ reveals a
structured distribution in the (wave-number, frequency) space which
is mostly concentrated on the curve defined by the dispersion
relation (\ref{HI.disp}). Therefore the strongly chaotic motion in
the fluid-sliding state is mainly caused by weak nonlinear
interactions between phonons. 

Here we will give only some qualitative arguments for the occurrence
of the peaks beyond the main resonance. The argument is based on the
picture of multi-phonon scattering processes already used in 
subsection~\ref{HI.L} to explain parametric resonance. The most
general process which drives the phonon system is the decay of $m$
washboard waves into $n$ different phonons. From energy and momentum
conservation we get
\begin{mathletters}
\label{HI.decay}
\begin{equation}
  v=v_m^n\equiv\frac{\sum_{j=1}^n\omega(k_j)}{2\pi m}
  \label{HI.decay.v}
\end{equation}
and 
\begin{equation}
  2\pi(cm+u)=\sum_{j=1}^nk_j,\quad\mbox{with $u$ integer}.
  \label{HI.decay.c} 
\end{equation} 
\end{mathletters} 
The integer $u$ denotes possible Umklapp processes.  The strength of
the decay processes scales like $b^m$. The most important processes
are those with small $n$ and $m$.  Superharmonic resonance (i.e.,
$n=1$) and parametric resonance (i.e., $n=2$) are the linear cases.
For $n>2$ the decay processes are nonlinear. Beyond first-order
parametric resonance (i.e., $v>\max v_1^P$) the solid-sliding state
is therefore insensitive to these processes. They become active only
if at least some phonons are excited. Then they excite additional
phonons leading, to a self-sustained process that converts
macroscopic, uniform sliding into microscopic, irregular motion. 

An important quantity characterizing the efficiency of phonon driving
due to washboard waves is the density of combinations of wave vectors
fulfilling (\ref{HI.decay}) for sliding velocities from the interval
of $[v,v+dv]$. We call this density the {\em density of decay
channels\/}. We expect that characteristic structures of this
density, such as peaks, also appear in $F_K(v)$. 

Because of the constant wave-number density due to periodic boundary
conditions, the density of decay channels of an $n$-wave decay
process of order $m$ reads
\begin{equation}
  \rho_m^n(v)=\frac{1}{(2\pi)^{n-1}}\int\limits_0^{2\pi}\cdots
   \int\limits_0^{2\pi}\delta(v-v_m^n)\,dk_1\ldots dk_{n-1},
  \label{HI.rho}
\end{equation}
where $\delta$ is Dirac's delta function and $k_n$ in $v_m^n$ is
substituted due to (\ref{HI.decay.c}). 

\begin{table}[phtb]
\caption[extrema]{\protect\label{t.extrema}The extrema of the main
decay processes for $\kappa=1$, $m=1$, and (a) $c=(\sqrt{5}-1)/2$ and
(b) $c=1$. $p$ and $n-p$ are the number of phonons having wave number
$k_-$ and $k_+$, respectively.}
\squeezetable
(a)
\begin{tabular}{cccccc}
$n$&$v_m^n$&$p$&$k_-$&$k_+$&Type\\ \hline
2 & 0.4800 & 1 & 5.4049 & 4.7615 & minimum\\
2 & 0.4801 & 0 &        & 5.0832 & maximum\\
2 & 0.6143 & 0 &        & 1.9416 & maximum\\ \hline
3 & 0.6052 & 0 &        & 5.4832 & minimum\\
3 & 0.7475 & 1 & 0.9028 & 1.4902 & saddle\\
3 & 0.7480 & 0 &        & 1.2944 & maximum\\
3 & 1.0611 & 0 &        & 3.3888 & maximum\\ \hline
4 & 0.7395 & 0 &        & 5.6832 & minimum\\
4 & 0.8707 & 0 &        & 0.9708 & minimum\\
4 & 0.8734 & 1 & 1.7615 & 0.7072 & saddle\\
4 & 1.2191 & 1 & 6.1524 & 3.4324 & saddle\\
4 & 1.2936 & 0 &        & 4.1124 & maximum\\
4 & 1.3729 & 0 &        & 2.5416 & maximum\\ 
\end{tabular}
(b)
\begin{tabular}{cccccc}
$n$&$v_m^n$&$p$&$k_-$&$k_+$&Type\\ \hline
2 & 0.3183 & 0 &        & 0.0000 & minimum\\
2 & 0.7118 & 0 &        & 3.1416 & maximum\\ \hline
3 & 0.4775 & 0 &        & 0.0000 & minimum\\
3 & 0.8709 & 1 & 0.0000 & 3.1416 & saddle\\
3 & 0.9549 & 0 &        & 2.0944 & maximum\\ \hline
4 & 0.6366 & 0 &        & 0.0000 & minimum\\
4 & 1.0300 & 2 & 0.0000 & 3.1416 & saddle\\
4 & 1.0950 & 1 & 0.6283 & 1.8850 & saddle\\
4 & 1.1027 & 0 &        & 1.5708 & maximum\\
4 & 1.4235 & 0 &        & 3.1416 & maximum\\ 
\end{tabular}
\end{table}

Extrema of $v_m^n$ lead to singularities in $\rho_m^n$ which are
generalizations of van-Hove singularities well-known from phonon
spectra \cite{wei.65}. In appendix~\ref{AP4} we have calculated the
extrema. For the parameters of Fig.~\ref{f.fk} the most important
extrema are tabulated in table~\ref{t.extrema}.  The scaling behavior
of $\rho_m^n$ near an extremum depends on its type. The most
prominent singularities are the diverging ones.  There are two cases
of diverging singularities. The first one occurs only for $n=2$ where
the density scales like $|v-v_c|^{-1/2}$ near maxima and minima of
$v_m^n$. For $n>2$ logarithmic singularities occur near certain types
of saddles.  A saddle can be characterized by the numbers $n_+$ and
$n_-$ which count the linearly independent directions in which
$v_m^n$ increases or decreases, respectively.  Logarithmic
singularities occur only if both numbers are odd (see
appendix~\ref{AP3}).

For the parameters of Fig.~\ref{f.fk} we have calculated numerically
the density $\rho_m^n$ of the most important decay processes. The
results are shown in figure~\ref{f.den}. Together with
table~\ref{t.extrema} we are able to identify all peaks in $F_K(v)$
that occur beyond the main resonance.

Concerning the strength of friction this is only a qualitative
analysis. The reason for that is twofold. First, the decay channels
are not equally fed and, second, we do not know the effective damping
constant of the phonon modes due to nonlinear phonon scattering. This
information would be needed in order to obtain a quantitative theory
of $F_K(v)$.

\subsection{\protect\label{HI.S}Coherent motions and solitons}

As already mentioned in the previous subsection, the motion is not
entirely chaotic near the large peaks at $v=1.06$ and $v=0.95$ for
$c=144/233$ and $c=1$, respectively. Coherent waves occur, as can be
seen in Figure~\ref{f.coh}. A spatio-temporal Fourier analysis shows
a peak which does not sit on the dispersion relation but near the
point which is determined by the wave number and frequency of the
phonon of the decay process that belongs to the absolute maximum of
$v_1^3$. 

In the example shown in Fig.~\ref{f.coh} we observe a modulation of
the amplitude of the fast oscillation. We have extracted the
amplitude of this envelope from the data of the simulation.
Figure~\ref{f.soli} shows the spatio-temporal evolution of this
amplitude. The regular motion of the equidistant maxima of the
envelope in Fig.~\ref{f.coh} appears as parallel white stripes in
Fig.~\ref{f.soli}. On this background pattern three dark stripes are
visible. They are a kind of {\em dark solitons\/} well-known from the
nonlinear Schr\"odinger equation. Since they propagate with different
velocities, collisions occur which are nondestructive like in the
nonlinear Schr\"odinger equation. Thus the FKT model seems to behave
like a conservative, undamped and undriven system. But the waves
emitted by the solitons are strongly damped.

\section{\protect\label{CON}Conclusion}

We have studied the dynamical properties of the FKT model driven at a
constant sliding velocity $v$. The model describes in a rudimentary
way the interface between two atomically flat sliding surfaces. An
advantage of this model is that some basic questions concerning dry 
friction can be studied analytically. 

Since dry friction operates mostly far from equilibrium, we are in
principle not able to calculate macroscopic properties by using
straightforward recipes. In the sliding process of two solids, the
uniform motion is turned into microscopic, irregular motion called
heat. The rate of dissipation determines the kinetic friction $F_K$.
What are the mechanisms of this transfer from macroscopic to
microscopic motion? How effective are these mechanisms? 

The two basic dissipation channels in the FKT model are the damping 
terms which are proportional to the velocities of the interface
particles relative to the sliding bodies. The basic mechanisms of
energy transfer are multi-phonon decay processes where $m$ so-called
washboard waves excite $n$ phonons of the chain of interface
particles. The washboard is the spatially periodic potential
describing the atomically flat surface of the lower body. It acts as
a wave (wave length~$=$ surface lattice constant, frequency~$=$
sliding velocity divided by wave length) in the frame of the upper
body to which the surface particles are connected by springs. The
washboard is driving the surface particles. For small driving
strength $b$ and large damping the process with $m=1$ and $n=1$ is
the most dominant one. It simply leads to a resonance peak in the
kinetic friction $F_K(v)$ given by (\ref{HI.FKmr}). A broad peak is
exactly what is found in boundary lubrication \cite{yos.93}. For
large velocities the washboard shakes the particles so rapidly that
its influence is averaged out. Therefore the interaction becomes
effectively zero. The interface motion freezes out forming the
solid-sliding state. In this state the kinetic friction is
proportional to the sliding velocity. Note that in Figs.~\ref{f.res}
and \ref{f.fk} this part of the kinetic friction is always omitted.

For increasing $b$ or decreasing damping superharmonic resonance
peaks occur. They are caused by processes with $m>1$ and $n=1$. The
superharmonic resonance peaks are smaller than the main resonance
peak and occur for velocities below the main resonance. As long as
the oscillation amplitudes of the interface particles are much less
than the washboard wave-length, the main resonance and superharmonic
resonances can by approximated analytically very well from the
linearized equation of motion.  Otherwise processes with $n>1$ become
important leading to nonlinear phonon damping. This increases the
effective damping constant resulting in a decrease of the resonance
peaks in accordance with (\ref{HI.FKmr.max}).

The decay process with $m=1$ and $n=2$ is responsible for an
instability (first-order parametric resonance) of the solid-sliding
state if $b$ exceeds some threshold which is proportional to the
damping constant. This instability occurs for a whole interval of $v$
beyond main resonance. The amplitudes of both phonons increase.
Nonlinearities in the equation of motion eventually saturate the
growth due to nonlinear phonon damping. This leads to quasiperiodic
motion because the phonon frequencies are in general incommensurate.
The kinetic friction in the parametric resonance window can be larger
than at the main resonance. 

A further increase of $b$ leads to bistability between the
solid-sliding state and the so-called fluid-sliding state which is
characterized by spatio-temporal chaos. The kinetic friction of this
state exhibits some peaks which can be qualitatively understood by
the density of decay channels for processes with $m=1$ and $n=2$,
$3$, and $4$. In the regime of fluid-sliding states a small velocity
interval is embedded where coherent motion including nondestructive
collisions of dark envelope solitons occurs.

It should be noted that the behavior beyond main resonance does not
depend on the commensurability of the ratio $c$ of the surface
lattice constants of the sliding bodies, contrary to what is found
for small sliding velocities.

The behavior for small sliding velocities is dominated by stick-slip
motion of the interface particles. In the overdamped limit every
$Q$-th particle ($Q$ is the denominator of $c$) jumps into the 
neighboring potential well leading to $F_S-F_K(v\to 0)\sim 1/Q$. Thus
the kinetic friction $F_K(v\to 0)$ is identical with the static
friction $F_S$ in the incommensurate case. The kinetic friction is
proportional to the sliding velocity if $b<b_c^K$. In any case
$F_K(v)$ is an increasing function at $v=0$. For finite damping
constants particles may jump to more distant potential wells leading
to kinetic friction which is less than the kinetic friction in the
overdamped limit. Furthermore, this stick-slip motion becomes more
irregular.

In order to compare the dynamical behavior of the FKT model with the
BK model we have to note that most studies at the BK model are done
for $\kappa\ll 1$ whereas in our study we have always chosen
$\kappa=1$. The behavior of the BK model is different for $\kappa=1$
compared to $\kappa\ll 1$. For example, the solitonic motion of a
group of a few blocks found by Schmittbuhl {\em et al.\/}
\cite{sch.93} disappears if $\kappa$ is too large \cite{esp.94}.  For
$\kappa=1.2$ Sarkadei and Jacobs \cite{sar.95} have found regular
periodic waves. They travel from the open end into the chain and
annihilate somewhere in the middle. But contrary to the FKT model,
this behavior does not lead to resonance peaks in the kinetic
friction \cite{esp.94,sar.95}. The reason for that may be the fact
that in the BK model energy is dissipated during the sticking phases
of the blocks and not during sliding.

There are several theoretical studies
\cite{per.96,str.96,per.93,mat.94,gra.96,bra.97} where the kinetic friction is
investigated in a system which is driven by a constant force and not 
by a constant velocity as in our study of the FKT model. The averaged
velocity of the center of mass is calculated. All these studies show
fluid-sliding states with enhanced friction and solid-sliding states
where the kinetic friction is given by the sliding velocity times the
damping constant. Hysteresis between the solid-sliding state and fluid-sliding
state occurs only if the temperature is small enough \cite{bra.97}.
Therefore we expect that the strong hysteresis in the FKT model 
disappears for finite and large enough temperatures.

A model which goes beyond the FKT model has been studied by Sokoloff
\cite{sok.84,sok.90,sok.92}. Instead of one layer it has several
layers coupled harmonically. Thus the driving due to the washboard
wave is no longer uniform. Sokoloff has calculated the kinetic
friction analytically for a prescribed periodic motion of the layer 
which feels the periodic potential. The result is therefore similar
our results for the main resonance and superharmonic resonance 
(if the prescribed motion has higher harmonics). This kind of
approach is not able to deal with instabilities like parametric
resonances or other multi-phonon processes. It would be worthwhile to
study such an extension of the FKT model from our point of view.

In a forthcoming paper we will study macroscopic stick-slip motion of
the FKT model by coupling the upper sliding body via a spring at a
support which moves with constant velocity. We are interested in the
following questions. What is the interplay between static friction 
and kinetic friction? What happens during the transitions from stick
to slip and vice versa? Which (meta)stable states does the system
select after the slip-to-stick transition? 

\section*{Acknowledgments}

We gratefully acknowledge helpful discussions with S. Aubry, T.
Baumberger, T.  Gyalog, M.~O. Robbins, R. Schilling, T. Strunz, and
H. Thomas.  We especially thank H. Thomas for careful reading of the
manuscript. We are grateful for the possibility to do simulations on
the NEC SX-3 at the Centro Svizzero di Calcolo Scientifico at Manno,
Switzerland. This work was supported by the Swiss National Science
Foundation.

\appendix

\section{\label{AP1}Quasistatic sliding in the Tomlinson limit and 
   for \protect\lowercase{$c=1$}}

In the Tomlinson limit (i.e., $\kappa,b\to\infty$) and for $c=1$ the
static hull function $g_S$ can be given in parametric form because
the next-neighbor coupling in (\ref{DEF.eqm2}) can be dropped:
\cite{wei.96a}
\begin{equation}
  g_S(\tau)=\frac{b}{\kappa}\sin 2\pi\tau,\quad x(\tau)=
   \tau-g_S(\tau).
  \label{AP1.gs}
\end{equation}
In order to calculate $F_K$ in the quasistatic limit $v\to 0$ we have
to distinguish between $b<b_c^K=\kappa/(2\pi)$ and $b>b_c^K$. 

In the first case we can apply (\ref{LOW.FK1}) with $\gamma_{\rm
eff}$ defined by (\ref{LOW.gamma}). To do this the integral
$\int_0^1(dg_S/dx)^2dx$ has to be calculated. Using (\ref{AP1.gs}) we
get
\begin{equation}
  \int_0^1\frac{(dg_S/d\tau)^2}{1-dg_S/d\tau}\,d\tau=
   \frac{1}{\sqrt{1-\left(\frac{2\pi b}{\kappa}\right)^2}}-1.
\end{equation}
Thus
\begin{equation}
  \gamma_{\rm eff}=\gamma_L+\gamma\left(\frac{1}{\sqrt{1-\left(
   \frac{2\pi b}{\kappa}\right)^2}}-1\right).
  \label{AP1.gamma}
\end{equation}

For $b>b_c^K$ the hull function $g_S$ is no longer a unique function
of $x$. There is a saddle-node bifurcation at the relative maximum of
$x(\tau)$ at 
\begin{equation}
  2\pi\tau_{SN}=\arccos\frac{\kappa}{2\pi b}.
  \label{AP1.tauSN}
\end{equation}
For the same value of $x$ there is at least a second solution
defining a $\tau_2$ implicitly by
\begin{equation}
  \tau_2-g_S(\tau_2)=\tau_{SN}-g_S(\tau_{SN}).
  \label{AP1.tau2}
\end{equation}
This equation can be solved perturbatively near $b=b_c^K$ and
$\kappa/b\to 0$ leading to
\begin{eqnarray}
  2\pi\tau_2&=&4\sqrt{\frac{\pi}{\kappa}(b-b_c^K)}-\frac{38}{15}
   \left(\frac{\pi}{\kappa}(b-b_c^K)\right)^{3/2}+\nonumber\\
   &&+\frac{1037}{350}\left(\frac{\pi}{\kappa}(b-b_c^K)\right)^{5/2}+
   {\cal O}\bigl((b-b_c^K)^{7/2}\bigr)
  \label{AP1.tau2.c}
\end{eqnarray}
and
\begin{eqnarray}
  2\pi\tau_2&=&2\pi n-\frac{\pi}{2}-\sqrt{\frac{2n\kappa}{b}}
   +\frac{\kappa}{2\pi b}-\frac{1}{3}\left(\frac{n\kappa}{2b}
   \right)^{3/2}+\nonumber\\&&+\frac{n}{6\pi}\left(\frac{\kappa}
   {b}\right)^2+{\cal O}\bigl((\kappa/b)^{5/2}\bigr),
  \label{AP1.tau2.infty}
\end{eqnarray}
with $n=1,2,\ldots$, respectively.

Because $x$ corresponds to the relative position of the sliding
bodies, an increase of $x$ beyond the relative maximum yields a jump
of one particle into a state corresponding to the solution of
(\ref{AP1.tau2}). In order to calculate the kinetic friction it is
most convenient to investigate the difference of the potential energy
before and after the jump:
\begin{eqnarray}
  \Delta V&=&\frac{b}{2\pi}[\cos 2\pi\tau_{SN}-\cos 2\pi\tau_2]
   +\nonumber\\&&+\frac{\kappa}{2}[g_S(\tau_{SN})^2-g_S(\tau_2)].
  \label{AP1.DV}
\end{eqnarray}
This expression can be simplified by using (\ref{AP1.gs}) and 
(\ref{AP1.tauSN}):
\begin{equation}
  \Delta V=\frac{\kappa}{8\pi^2}\left(1-\frac{2\pi b}{\kappa}
      \cos 2\pi\tau_2\right)^2.
  \label{AP1.DV2}
\end{equation}
In accordance with (\ref{LOW.FK0}) the kinetic friction reads
\begin{equation}
  F_K(v\to 0)=\frac{\Delta V}{\langle n\rangle}\,N,
  \label{AP1.FK}
\end{equation}
where $\langle n\rangle$ is the number of potential wells a particle
jumps on average. For $b\to b_c^K$ a particle can jump only into the
next potential well.  Thus $\langle n\rangle=1$ and we get with
(\ref{AP1.tau2.c}) 
\begin{eqnarray}
  \frac{F_K(v\to 0)}{N}&=&\frac{9}{2\kappa}(b-b_c^K)^2-\frac{36\pi}
   {5\kappa^2}(b-b_c^K)^3+\nonumber\\&&+\frac{2124\pi^2}{175\kappa^3}
   (b-b_c^K)^4+{\cal O}\bigl((b-b_c^K)^5\bigr).
  \label{AP1.FK.c}
\end{eqnarray}
In the limit $\kappa/b\to 0$ we get with (\ref{AP1.tau2.infty})
\begin{eqnarray}
  \frac{F_K(v\to 0)}{N}&=&b\left(1-\frac{\langle n\rangle\kappa}{2b}
   +\right.\nonumber\\&&+\left.\frac{\sqrt{2\langle n\rangle}}{3\pi}
   \left(\frac{\kappa}{b}\right)^{3/2}+{\cal O}\bigl((\kappa/b)^2
   \bigr)\right).
  \label{AP1.FK.infty}
\end{eqnarray}
These equations generalize the results of Ref.~\onlinecite{hel.94}.

\section{\label{AP2}Perturbation theory for the linearized equation 
   of motion}

The linearized equation of motion (\ref{HI.leqm}) can be treated
analytically in the framework of a perturbation theory with $b$
as the smallness parameter. 

First the solution is calculated in the following way. We start with 
(\ref{HI.Am}) and expand the solution:
\begin{equation}
  A_m=\sum_{l=1}^\infty A_{m,l}b^l.
  \label{AP2.Am}
\end{equation}
Inserting this ansatz into (\ref{HI.Am}) leads to a recursion
equation for the coefficients $A_{m,l}$:
\begin{equation}
  A_{m,1}=\frac{i}{2L_m}(\delta_{m,-1}-\delta_{m,1})
  \label{AP2.Am1}
\end{equation}
and
\begin{equation}
  A_{m,l+1}=\frac{\pi}{L_m}(A_{m+1,l}+A_{m-1,l}).
  \label{AP2.Aml.re}
\end{equation}
The solution of this recursion scheme is
\begin{equation}
 A_{m,l}=-\frac{i\pi^{l-1}}{2}
   \sum_{\sigma_1,\ldots,\sigma_l=-1\atop\sum_{i=1}^l\sigma_i=m}^1
   \sigma_1\prod_{i=1}^l\frac{1}{L_{\sum_{j=1}^i\sigma_j}}.
  \label{AP2.Aml}
\end{equation}
Thus the only nonzero coefficients are
\begin{equation}
  A_{m,m+2n},\quad{\rm with}\quad m+n\not=0,\ m,n=0,1,\ldots
  \label{AP2.Aml0}
\end{equation}
Therefore the leading term of the Fourier component $A_m$ is of order
$b^m$:
\begin{eqnarray}
  A_m&=&-\frac{i(\pi b)^m}{2\pi}\left(\prod_{i=1}^m\frac{1}{L_i}
   \right)\times\nonumber\\&&\times\left(1-\frac{(\pi b)^2}{L_{-1}
   L_0}+\sum_{k=0}^m\frac{(\pi b)^2}{L_kL_{k+1}}+{\cal O}(b^4)\right)
  \label{AP2.Amm}
\end{eqnarray}
Using (\ref{HI.FKF}) we get for the kinetic friction
\begin{equation}
   \frac{F_K}{N}=8\pi^2v\gamma\sum_{l=1}^\infty b^{2l}\sum_{m=1}^lm^2
   \sum_{n=0}^{l-m}A_{m,m+2n}A^*_{m,2l-m-2n},
  \label{AP2.FK}
\end{equation}
which is an even function in $b$ as expected from symmetry.

In order to study the stability of the solutions of (\ref{HI.leqm}), 
we make the following ansatz in accordance with the well-known 
Floquet-Bloch theorem :
\begin{equation}
  \xi_j=\sum_{m=-\infty}^\infty a_me^{2\pi im(vt+cj)+iqj+\lambda t},
  \label{AP2.xi}
\end{equation}
where $q$ is an arbitrary given wave number (with $qN$ an integer
multiple of $2\pi$) and $\lambda$ a complex number which is an
eigenvalue of:
\begin{equation}
  D_ma_m-\pi b(a_{m-1}+a_{m+1})=0,
  \label{AP2.evp}
\end{equation}
with
\begin{equation}
  D_m=(2\pi ivm+\lambda)^2+\gamma(2\pi ivm+\lambda)
   +\omega^2(2\pi cm+q).
  \label{AP2.Dm}
\end{equation}
Note that (\ref{AP2.evp}) is invariant under the transformation
$\lambda\to\lambda+2\pi iv$, $q\to q+2\pi c$, $a_m\to a_{m+1}$.  A
solution of (\ref{HI.leqm}) is stable as long as the real parts of
all eigenvalues are negative. In order to calculate the instability
threshold $b_c$ it is more convenient to solve (\ref{AP2.evp}) for
$\lambda=i\Omega$ where $b$ and $\Omega$ are the unknown variables.
The diagonal elements turn into
\begin{equation}
  D_m=R_m+i\gamma I_m,
  \label{AP2.Dm0}
\end{equation}
with
\begin{equation}
  R_m=\omega^2(2\pi cm+q)-I_m^2,\quad I_m=2\pi vm+\Omega.
  \label{AP2.RmIm}
\end{equation}
The real solution $b$ of (\ref{AP2.evp}) with the smallest absolute
value defines a function of $q$. Its absolute minimum gives the
instability threshold $b_c$. It is always zero in the undamped case
(i.e., $\gamma=0$) because at least one $D_m$ is zero. In the case of
damping all $D_m$'s are nonzero.

Parametric resonance of order $n$ corresponds to the fact that two
diagonal elements, say $D_0$ and $D_n$, go to zero for $b\to 0$ and 
$\gamma\to 0$. In order to calculate the instability threshold we
assume that $a_0$ and $a_n$ are given. Next we express $a_{\pm 1}$
and $a_{n\pm 1}$ in terms of $a_0$ and $a_n$:
\begin{equation}
  a_{-1}=-\frac{\pi b\,d_{-\infty}^{-2}}{d_{-\infty}^{-1}}a_0,
  \label{AP2.am1}
\end{equation}
\begin{equation}
  a_1=-\frac{\pi b\,d_2^{n-1}}{d_1^{n-1}}a_0-\frac{(-1)^n(\pi b)^
   {n-1}}{d_1^{n-1}}a_n,
  \label{AP2.a1}
\end{equation}
\begin{equation}
  a_{n-1}=-\frac{\pi b\,d_1^{n-2}}{d_1^{n-1}}a_n-\frac{(-1)^n(\pi b)^
   {n-1}}{d_1^{n-1}}a_0,
  \label{AP2.anm1}
\end{equation}
\begin{equation}
  a_{n+1}=-\frac{\pi b\,d_{n+2}^\infty}{d_{n+1}^\infty}a_n,
  \label{AP2.amp1}
\end{equation}
where
\begin{equation}
  d_n^m=\left|\begin{array}{cccc}D_n&\pi b&&0\\\pi b&D_{n+1}&\ddots\\
    &\ddots&\ddots&\pi b\\0&&\pi b&D_m\end{array}\right|.
  \label{AP2.detnm}
\end{equation}
Eq.~(\ref{AP2.evp}) for $m=0$ and $m=n$ turns into a set of linear
equations for $a_0$ and $a_n$. It has nontrivial solutions if the
determinant
\begin{equation}
  \left|\begin{array}{cc}
   \frac{\displaystyle D_0}{\displaystyle(\pi b)^2}-\frac{
   \displaystyle d_{-\infty}^{-2}}{\displaystyle d_{-\infty}^{-1}}
   -\frac{\displaystyle d_2^{n-1}}{\displaystyle d_1^{n-1}}&
   -(-1)^n\frac{\displaystyle(\pi b)^{n-2}}{\displaystyle d_1^{n-1}}
   \\[2mm]-(-1)^n\frac{\displaystyle(\pi b)^{n-2}}{\displaystyle 
   d_1^{n-1}}&\frac{\displaystyle D_n}{\displaystyle(\pi b)^2}-
   \frac{\displaystyle d_{n+2}^\infty}{\displaystyle d_{n+1}^\infty}
   -\frac{\displaystyle d_1^{n-2}}{\displaystyle d_1^{n-1}}
   \end{array}\right|
  \label{AP2.det}
\end{equation}
is zero. The assumption 
\begin{equation}
  R_0,R_n={\cal O}(b^2),\quad R_m={\cal O}(b^0),\ m\not=0,n,
  \label{AP2.scale}
\end{equation}
yields
\begin{equation}
  \left(\prod_{m=1}^{n-1}D_j+{\cal O}(b^2)\right)^2\left[-\gamma^2I_0
   I_n+{\cal O}(b^2)\right]=(\pi b)^{2n}.
  \label{AP2.det0}
\end{equation}
Thus in leading order of $\gamma$ we obtain
\begin{equation}
  \pi b_c=\left(\gamma\sqrt{-I_0I_n}\prod_{m=1}^{n-1}|R_j|\right)
   ^{1/n}.
  \label{AP2.bc}
\end{equation}
The product term does not appear for $n=1$. Bearing in mind that
$I_0I_n$ has to be negative and $\omega(k)$ is an even function we
immediately get the parametric resonance condition (\ref{HI.pr.vn})
by solving $R_0=R_n=0$.

\section{\label{AP4}Extrema of \protect\lowercase{$v_m^n$}}

Before calculating the extrema of $v_m^n$ defined by (\ref{HI.decay})
we eliminate $k_n$ by using (\ref{HI.decay.c}). Thus $v_m^n$ is a
function of $k_1,\ldots,k_{n-1}$. An extremum is given by
\begin{equation}
  2\pi m\partial_{k_j}v_m^n=\omega'(k_j)-\omega'(k_n)=0,\quad
   j=1,\ldots,n-1,
  \label{AP4.dv}
\end{equation}
where
\begin{equation}
  \omega'(k)=\frac{d\omega}{dk}=\frac{\sin k}{\omega(k)}
  \label{AP4.ws}
\end{equation}
is the derivative of the dispersion relation (\ref{HI.disp}). For an
extremum all $\omega'(k_j)$ have to be the same. Because $\omega'(k)$
is a periodic function, there exist two solutions $k_-$ and $k_+$
in the interval $[-\pi,\pi)$. For a given value of $k_+$ we can
calculate the corresponding $k_-$ by solving
$\omega'(k_-)=\omega'(k_+)$:
\begin{equation}
  k_-=\mbox{sign}\,(k_+)\arccos\left(\frac{2-(\kappa+2)\cos k_+}
   {\kappa+2-2\cos k_+}\right),
  \label{AP4.kp}
\end{equation}
where sign is the signum function. Therefore an extremum 
is given by
\begin{equation}
  v_m^n=\frac{p\omega(k_-)+(n-p)\omega(k_+)}{2\pi m},
  \label{AP4.v}
\end{equation}
where $p$ and $n-p$ are the numbers of wave vectors which have the
wave numbers $k_-$ and $k_+$, respectively. 

The wave number $k_-$ is not arbitrary because momentum conservation 
(\ref{HI.decay.c}) has to be satisfied:
\begin{equation}
  pk_-+(n-p)k_+=2\pi(mc+u),\quad\mbox{with $u$ integer}.
  \label{AP4.ksum}
\end{equation}
Eliminating $k_-$ in 
(\ref{AP4.ksum}) by using (\ref{AP4.kp}) leads to a nonlinear
algebraic equation for $k_+$ which can be solved only numerically
except for $p=0$. In this case all wave vectors are identical. That
is,
\begin{equation}
  k_j=k_+=2\pi\frac{mc+u}{n},\quad j=1,\ldots,n.
  \label{AP4.k.p0}
\end{equation}

The type of the extremum is determined by the eigenvalues of 
\begin{equation}
  2\pi m\partial_{k_j}\partial_{k_l}v_m^n=\omega''(k_j)\delta_{j,l}
   +\omega''(k_n),
  \label{AP4.ddv}
\end{equation}
where $\omega''(k)$ is the second derivative of the dispersion
relation (\ref{HI.disp}). For $p=0$ the eigenvalues are
\begin{equation}
  \lambda_1=n\omega''(k_+),\quad \lambda_j=\omega''(k_+),\quad
   j=2,\ldots,n-1.   
  \label{AP4.l.p0}
\end{equation}
Thus the extremum is either a maximum or minimum depending on the
sign of $\omega''$. In the general case we have two eigenvalues
given by
\begin{equation}
  \lambda=\frac{\omega''_-+n\omega''_+\pm\sqrt{(\omega''_--
  n\omega''_+)^2+4p\omega''_+(\omega''_--\omega''_+)}}{2},
  \label{AP3.l.p}
\end{equation}
where
\begin{equation}
  \omega''_\pm\equiv\omega''(k_\pm),
  \label{AP4.wsspm}
\end{equation}
$p-1$ times the eigenvalue $\lambda=\omega''_-$, and $n-p-2$ times 
the eigenvalue $\lambda=\omega''_+$. 

\section{\label{AP3}Scaling of \protect\lowercase{$\rho_m^n$} near 
   extrema of \protect\lowercase{$v_m^n$}}

The scaling behavior of $\rho_m^n$ near an extremum of
$v_m^n(k_1,\ldots,k_{n-1})$ can be easily obtained after a linear
transformation of the coordinate system $k_1,\ldots,k_{n-1}$ into a
new system $x_1,\ldots,x_{n-1}$ which is given by the eigendirections
of the matrix defined by the second derivatives of $v_m^n$ calculated
at the extremum. Up to second order we get
\begin{equation}
  v_m^n=v_c+\sum_{j=1}^{n_+}x_j^2-\sum_{j=1}^{n_-}x_j^2,
  \label{AP3.v}
\end{equation}
where $n_+$ ($n_-$) is the number of positive (negative) eigenvalues
of that matrix. A non-singular matrix leads to $n_++n_-=n-1$,
otherwise $n_++n_-<n-1$.

The scaling behavior near an extremum is determined by the states
nearby.  Thus we perform the integration (\ref{HI.rho}) only in a
spherical domain of radius $R$:
\begin{equation}
  \rho_m^n\sim\int_0^R\int_0^R\delta(v-v_c-r_+^2+r_-^2)r_+^{n_+-1}
   r_-^{n_--1}dr_+dr_-,
  \label{AP3.rho}
\end{equation}
where $r_\pm^2\equiv\sum_{j=1}^{n_\pm}x^2_j$. One integration yields
\begin{equation}
  \rho_m^n\sim\int_0^R(v-v_c-r_-^2)^{n_+/2-1}r_-^{n_--1}dr_-.
  \label{AP3.rho2}
\end{equation}
In the case of either a relative minimum or maximum (i.e., $n_-=0$ 
or $n_+=0$) we get therefore
\begin{equation}
  \rho_m^n\sim(v-v_c)^{n_+/2-1}\quad{\rm or}\quad
  \rho_m^n\sim(v_c-v)^{n_-/2-1}.
  \label{AP3.rs1}
\end{equation}
In the case of a saddle (i.e., $n_\pm\not=0$) we perform the
second integration (\ref{AP3.rho2}) by using the Euler substitution 
\begin{displaymath}
  t=r_-+\sqrt{z+r_-^2},\quad{\rm with}\quad z\equiv v-v_c,
\end{displaymath}
which yields
\begin{equation}
  \rho_m^n\sim\int\limits_{\sqrt{z}}^{\sqrt{z+R^2}+R}
   (t^2+z)^{n_+-1}(t^2-z)^{n_--1}t^{1-n_+-n_-}dt.
  \label{AP3.rho3}
\end{equation}
Neglecting the contribution from the upper limit we get
\begin{eqnarray}
  \rho_m^n&\sim&\sum\limits_{j_+=0}^{n_+-1}\sum\limits_{j_-=0}
   ^{n_--1}(-1)^{j_-}\left(n_+-1\atop j_+\right)\left(n_--1\atop 
   j_-\right)\times\nonumber\\&&\times z^{\frac{n_++n_-}{2}-1}
   I(n_+-n_--2-2j_+-2j_-),
  \label{AP3.rho4}
\end{eqnarray}
where
\begin{equation}
  I(s)\equiv\left\{\begin{array}{ll}\ln z,&s=0\\1,&s\not=0
   \end{array}\right..
  \label{AP3.I}
\end{equation}
The logarithmic term appears only if 
\begin{equation}
  C\equiv\sum_{j_\pm=0\atop 2(j_++j_-+1)=n_++n_-}^{n_\pm-1}
  (-1)^{j_-}\left(n_+-1\atop j_+\right)\left(n_--1\atop j_-\right)
  \label{AP3.C}
\end{equation}
is not zero. It is easy to show that $C=-(-1)^{n_\pm}C$ holds.
Thus
\begin{equation}
  \rho_m^n\sim|v-v_c|^{\frac{n_++n_-}{2}-1}\left\{\begin{array}{ll}
   \ln|v-v_c|,&n_\pm\ \mbox{are odd}\\1,&{\rm otherwise}
   \end{array}\right..
  \label{AP3.rs}
\end{equation}

\begin{figure}
\caption[f1]{\protect\label{f1}A schematical sketch of the
Frenkel-Kontorova-Tomlinson (FKT) model.
}
\end{figure}

\begin{figure}
\caption[flow]{\protect\label{f.low}The friction force in the limit
of quasistatic sliding (i.e., $v\to 0$) for $\kappa=1$. The solid
lines show the kinetic friction $F_K$ in the overdamped limit (i.e.,
$\gamma\to\infty$). The dashed lines denote the static friction
$F_S$. $F_S^{max}=bN$ is the overall upper bound of static friction.
For $c=1$ the analytically obtained approximations of $F_K$, 
(\ref{AP1.FK.c}) and (\ref{LOW.FK0.infty}), are shown by dotted
lines.
}
\end{figure}

\begin{figure}
\caption[flows]{\protect\label{f.lows}The kinetic friction $F_K$ for
small sliding velocities. Parameters are $\kappa=1$, $N=233$,
$M=144$, $b=2/\pi\approx 0.637$. Using definition (\ref{DEF.FK1}) the
data has been obtained by numerical integration of (\ref{DEF.eqm2}).
Each data point is the average over $500/v$ time units. In order to
avoid averaging during transients the system has been evolved over
$10^3$ time units before starting averaging. The last state has
been chosen as the initial value for the next data point with
increased $v$. In order to get some information about the
low-frequency fluctuations the averaging interval is chunked into
ten pieces, each $50/v$ time units long, yielding ten values of
$F_K$. The standard deviation of them are denoted by error bars.
}
\end{figure}

\begin{figure}
\caption[fres]{\protect\label{f.res}Main resonance, superharmonic and
parametric resonance for $\kappa=1$, $N=233$, and $M=144$. Resonance 
velocities of main resonance and superharmonic resonances in
accordance with (\ref{HI.vms}): $v^S_1\approx 0.337$, $v^S_2\approx
0.134$, $v^S_3\approx 0.071$, $v^S_4\approx 0.089$.  Parametric
resonance intervals in accordance with (\ref{HI.pr.vn}):  First
order:  $[0.480,0.614]$. Second order: $[0.197,0.337]$. Part (a)
shows a comparison between the approximation (\ref{HI.FKapp}) (dotted
line) and the numerically obtained solution of equation of the
dynamical hull function (\ref{HI.heqm}) (solid line) for $\gamma=0.5$
and $b=1/\pi\approx 0.318$. Part (b) shows different solutions of
(\ref{HI.heqm}) for $\gamma=0.1$ and $2\pi b=0.1$, $0.2$, $0.5$, $1$
and $2$. Solid (dotted) lines indicate stable (unstable) solutions.
Part (c) compares the hull function solution (dashed line) with the
full simulation of (\ref{DEF.eqm2}) (solid line and squares for
increasing $v$, dotted line and triangles for decreasing $v$) for
$\gamma=0.1$ and $b=1/(2\pi)\approx 0.159$.  Details of the
simulation are the same as in Fig.~\ref{f.lows}.
}
\end{figure}

\begin{figure}
\caption[fquasi]{\protect\label{f.quasi}Stroboscopic map of 
quasiperiodic and chaotic motion caused by first-order parametric
resonance. The parameters are $\kappa=1$, $N=233$, $M=144$,
$\gamma=0.1$, $v=0.5$, and (a) $b=1/(2\pi)\approx 0.159$, (b)
$b=3/(2\pi)\approx 0.478$. $f_n$ is the lateral force
$-\kappa/N\sum_j\xi_j$ at time $t=n/v$.
}
\end{figure}

\begin{figure}
\caption[ffk]{\protect\label{f.fk}The kinetic friction $F_K(v)$.
Parameters are $\kappa=1$, $N=233$, $b=3/(2\pi)\approx 0.478$,
$\gamma=0.1$, and (a) $c=144/233$, (b) $c=1$. Squares (triangles)
connected by solid (dashed) lines denote the numerically obtained
values for increasing (decreasing) sliding velocity $v$. The
averaging procedure is the same as described in the caption of
Fig.~\ref{f.lows}.
}
\end{figure}

\begin{figure}
\caption[fden]{\protect\label{f.den}Densities of the most important
decay channels beyond the main resonance. The parameters are
$\kappa=1$ and (a) $c=(\sqrt{5}-1)/2 \approx 0.618$, (b) $c=1$.
}
\end{figure}

\begin{figure}
\caption[fcoh]{\protect\label{f.coh}Coherent motion. The first five
particles are shown. The parameters are $\kappa=1$, $N=233$, $M=144$,
$\gamma=0.1$, $b=3/(2\pi)\approx 0.478$, and $v=1.06$.
}
\end{figure}

\begin{figure}
\caption[fsoli]{\protect\label{f.soli}Dark envelope solitons embedded
into the coherent motion seen in Fig.~\ref{f.coh}. The parameters are
the same as in Fig.~\ref{f.coh}. The grey scale from dark to bright
denotes the strength of the amplitude of the fast oscillations of the
particles. The coherent motion is indicated by the regular stripes.
There are three dark solitons traveling with three different
velocities. At $j\approx 100$ and $t\approx 190$ two of them interact
nondestructively.
}
\end{figure}


\begin{references}
\bibitem[*]{MWadr}Present Address: Institut f\"ur Angewandet Physik,
Universit\"at Bern, Sidlerstr. 5, CH-3012 Bern, Switzerland.
\bibitem{bow.54}
F.~P. Bowden and D. Tabor, {\it Friction and Lubrication} (Oxford
University Press, 1954).

\bibitem{sin.92}
I.~L. Singer and H.~M. Pollock (eds), {\it Fundamentals of friction:
macroscopic and microscopic processes} (Kluwer Academic Publishers,
Dordrecht, 1992).

\bibitem{hes.94}
F. Heslot, T. Baumberger, B. Perrin, B. Caroli, and C. Caroli, Phys.
Rev. E {\bf 49}, 4973 (1994).

\bibitem{bhu.95}
B. Bhushan, J.~N. Israelachvili, and U. Landman, Nature {\bf 374},
607 (1995).

\bibitem{wei.96a}
M. Weiss and F.~J. Elmer, Phys. Rev. B {\bf 53}, 7539 (1996).

\bibitem{tom.29}
G.~A. Tomlinson, Phil. Mag. Series 7, {\bf 7}, 905 (1929).

\bibitem{mcc.89}
G.~M. McClelland, in {\it Adhesion and Friction}, M. Grunze and H.J.
Kreuzer (eds.), Springer Series in Surface Science {\bf 17}, 1
(Springer Verlag, Berlin, 1990).

\bibitem{mcc.92}
G.~M. McClelland and J.~N. Glosli, in Ref.~\onlinecite{sin.92}, p. 
405.

\bibitem{sok.84}
J.~B. Sokoloff, Surf. Sci. {\bf 144}, 267 (1984).

\bibitem{hel.94}
J.~S. Helman, W. Baltensberger, and J.~A. Ho{\l}yst, Phys. Rev. B 
{\bf 49}, 3831 (1994).

\bibitem{elm.94}
F.~J. Elmer, Helv. Phys. Acta {\bf 67}, 213 (1994).

\bibitem{remINT.1}
In Ref.~\onlinecite{hel.94} also multiple resonances of the form 
$\omega=n\omega_0$, where $n$ is an integer greater than two, are
mentioned. But such resonances do not occur in the limit of
disappearing interaction and dissipation. Therefore we believe that
the occurrence of these resonances depends on the model.

\bibitem{remINT.2}
This is in contrast to the suggestion at the end of
Ref.~\onlinecite{hel.94}.

\bibitem{wei.96b}
M. Weiss and F.~J. Elmer, in {\em The physics of sliding friction\/},
B.~N.~J. Persson and E. Tosatti (eds.), (Kluwer Academic Publishers,
Dordrecht, 1996).

\bibitem{bur.67}
R. Burridge and L. Knopoff, Bull. Seismol. Soc. Am. {\bf 57}, 341 
(1967).

\bibitem{car.89}
J.~M. Carlson and J.~S. Langer, Phys. Rev. Lett. {\bf 62}, 2632 
(1989), Phys. Rev. A {\bf 40}, 6470 (1989).

\bibitem{car.91}
J.~M. Carlson, J.~S. Langer, B.~E. Shaw,  and C. Tang, Phys. Rev. A 
{\bf 44} 884 (1991).

\bibitem{vas.92}
G.~L. Vasconcelos, M. de Sousa Vieira, and S.~R. Nagel, Physica A 
{\bf 191}, 69 (1992).

\bibitem{des.93}
M. de Sousa Vieira, G.~L. Vasconcelos, and S.~R. Nagel, Phys. Rev. 
E {\bf 47}, R2221 (1993).

\bibitem{sch.93}
J. Schmittbuhl, J.-P. Vilotte, and S. Roux, Europhys. Lett. {\bf 21},
375 (1993).

\bibitem{des.94}
M. de Sousa Vieira and H.~J. Herrmann, Phys. Rev. E {\bf 49}, 4534
(1994).

\bibitem{esp.94}
P. Espa\~nol, Phys. Rev. E {\bf 50}, 227 (1994).

\bibitem{sar.95}
M.~R. Sarkadei and R.~L. Jacobs, Phys. Rev. E {\bf 51}, 1929 (1995).

\bibitem{des.96}
M. de Sousa Vieira, Phys. Rev. E. {\bf 54}, 5925 (1996).

\bibitem{hee.86}
D.W. Heermann, {\it Computer Simulation Methods in Theoretical
Physics}, Springer-Verlag (1986).

\bibitem{remLOW.1}
This should not be confused with the fact that for the {\em
undriven\/} FKT model, where the positions of the sliding bodies are
{\em not\/} externally controlled, metastable states occur not before
$b$ exceeds a second threshold $b_c^m\ge b_c^K$ (for more details,
see Ref.~\onlinecite{wei.96a}).

\bibitem{gya.94}
T. Gyalog, unpublished Diploma thesis, Universit\"at Basel.

\bibitem{elm.96}
F.~J. Elmer, in {\em The physics of sliding friction\/}, B.~N.~J.
Persson and E. Tosatti (eds.), (Kluwer Academic Publishers,
Dordrecht, 1996).

\bibitem{sok.90}
J.B. Sokoloff, Phys. Rev. B {\bf 42}, 760 (1990).

\bibitem{per.95}
B.~N.~J. Persson, J. Chem. Phys. {\bf 103}, 3849 (1995).

\bibitem{per.96}
B.~N.~J. Persson and A. Nitzan, Surf. Sci. {\bf 367}, 261 (1996).

\bibitem{remHI.1}
The first term of (\ref{DEF.FK2}) is always dropped throughout
Sec.~\ref{HI}.

\bibitem{yos.93}
H. Yoshizawa and J. Israelachvili, J. Phys. Chem. {\bf 97}, 11300
(1993).

\bibitem{ll.76}
L.D. Landau and L.M. Lifshitz, {\it Mechanics} (Pergamon Press,
Oxford, 1976).

\bibitem{han.65}
M. Abromowitz and I.A. Stegun (eds.), 
{\it Handbook of Mathematical Functions} (Dover, New York, 1965).

\bibitem{str.96}
T. Strunz and F.~J. Elmer, in {\em The physics of sliding 
friction\/}, B.~N.~J. Persson and E. Tosatti (eds.), (Kluwer 
Academic Publishers, Dordrecht, 1996).

\bibitem{cro.93}
M.~C. Cross and P.~C. Hohenberg, Rev. Mod. Phys. {\bf 65}, 851 
(1993).

\bibitem{wei.65}
G. Weinreich, {\it Solids: Elementary Theory for Advanced Students} 
(Wiley, New York, 1965).

\bibitem{per.93}
B.~N.~J. Persson, Phys. Rev. B {\bf 48}, 18140 (1993).

\bibitem{mat.94}
H. Matsukawa and H. Fukuyama, Phys. Rev. B {\bf 49}, 17286 (1994).

\bibitem{gra.96}
E. Granato, M.~R. Baldan, and S.~C. Ying, in {\em The physics of 
sliding friction\/}, B.~N.~J. Persson and E. Tosatti (eds.), 
(Kluwer Academic Publishers, Dordrecht, 1996).

\bibitem{bra.97}
O.~M. Braun, T. Dauxois, M.~V. Paliy, and M. Peyrard, preprint cond-mat/9701134.

\bibitem{sok.92}
J.B. Sokoloff, J. Appl. Phys. {\bf 72}, 1262 (1992).

\end{references}
\end{document}